\documentclass[conference]{IEEEtran}

\IEEEoverridecommandlockouts
\usepackage{cite}
\usepackage{amsmath,amssymb,amsfonts}
\usepackage{algorithmic}
\usepackage{graphicx}
\usepackage{textcomp}
\usepackage{xcolor}
\def\BibTeX{{\rm B\kern-.05em{\sc i\kern-.025em b}\kern-.08em
    T\kern-.1667em\lower.7ex\hbox{E}\kern-.125emX}}
\usepackage{graphicx}
\usepackage{subcaption}
\graphicspath{ {images/} }
\usepackage[utf8]{inputenc}
\usepackage{fancyhdr}
\usepackage{dirtytalk}
\usepackage{tabularx}
\PassOptionsToPackage{hyphens}{url}\usepackage{hyperref}
\usepackage[ruled,lined,linesnumbered]{algorithm2e}

\usepackage{enumitem,kantlipsum}

\usepackage{graphicx}
\graphicspath{ {images/} }
\usepackage[utf8]{inputenc}
\usepackage{fancyhdr}
\usepackage{dirtytalk}
\usepackage{tikz}
\usetikzlibrary{shapes.geometric, arrows}
\usepackage{pgfplots}
\pgfplotsset{width=8cm,compat=1.16}
\usepgfplotslibrary{external}
\usepackage{mathtools}

\newtheorem{theorem}{\normalfont{\textbf{Theorem}}}

\newtheorem{definition}{\normalfont{\textbf{Definition}}}

\tikzstyle{startstop} = [rectangle, rounded corners, minimum width=2cm, minimum height=0.5cm,text centered, draw=black, fill=red!30]
\tikzstyle{process} = [rectangle, minimum width=2cm, minimum height=0.5cm, text centered, draw=black, fill=orange!30, align=left]
\tikzstyle{decision} = [diamond, minimum width=1.0cm, minimum height=0.4cm, text centered, draw=black, fill=green!30]
\tikzstyle{arrow} = [thick,->,>=stealth]

\newcolumntype{L}{>{$}l<{$}}

\usepackage[labelformat=simple]{subcaption}

\usepackage[nolist]{acronym}
\begin{acronym}
\acro{ML}{machine learning}
\acro{AI}{artificial intelligence}
\acro{UE}{user equipment}
\acro{IB}{information bottleneck }
\acro{AE}{Auto-Encoder}
\acro{IC}{incentive compatibility}
\acro{IR}{individual rationality}
\acro{SSP}{spectrum service provider}
\end{acronym}

\begin{document}


\title{Auction-Driven Spectrum Allocation With AutoEncoder-Based Compression in Rural Wireless Networks: A Novel Framework for Reliable Telemedicine}







\author{\IEEEauthorblockN{Nadjemat El Houda Issaad\IEEEauthorrefmark{1},
Ismail Lotfi\IEEEauthorrefmark{2}, 
Mohamed Senouci\IEEEauthorrefmark{1}, and
Zekri Lougmiri\IEEEauthorrefmark{1}}
\IEEEauthorblockA{\IEEEauthorrefmark{1}Department of Computer Science, Ahmed Ben Bella Oran 1 University, Algeria.}

\IEEEauthorblockA{\IEEEauthorrefmark{2}College of Science and Engineering, Hamad Bin Khalifa University, Doha, Qatar.
}
}



\maketitle
\begin{abstract}
Rural healthcare faces numerous challenges, including limited access to specialized medical services and diagnostic equipment, which delays patient care. Enhancing the ability to transmit medical images and data from rural areas to urban hospitals via wireless networks is critical. However, bandwidth limitations, unreliable networks, and concerns over data security and privacy hinder efficient transmission. Additionally, the high data volume of medical content and the limited battery life of IoT devices pose further challenges. To address these challenges, data compression techniques such as Autoencoders (AEs) offer promising solutions by significantly reducing the communication overhead without sacrificing essential image quality or details. Additionally, spectrum allocation mechanisms in rural areas are often inefficient, leading to poor resource utilization. Auction theory presents a dynamic and adaptive approach to optimize spectrum allocation. This paper proposes a novel hybrid framework that integrates AE-based data compression with auction-based spectrum allocation, addressing both communication efficiency and spectrum utilization in rural wireless networks. Extensive simulations validate the framework’s ability to improve spectrum utilization, transmission efficiency, and overall connectivity, offering a practical solution for enhancing rural telemedicine infrastructure.

\end{abstract}




\section{Introduction}
\subsection{Background and Motivations}


Healthcare in rural areas faces major challenges, such as limited access to facilities and specialists, with patients often traveling long distances for care, causing delays. Local healthcare providers may lack advanced diagnostic tools and resources, making emergency response difficult. Enhancing communication between rural healthcare workers and main hospitals via mobile networks, particularly for transmitting medical images, is essential for timely care~\cite{Kasban_2021}.
Additionally, portable and wearable ultrasound devices are crucial in point-of-care diagnostics for cardiovascular, respiratory, and obstetrical conditions, offering real-time monitoring capabilities. These systems significantly enhance telemedicine by allowing physicians to remotely monitor patients with chronic conditions. Early detection of high-risk cardiovascular diseases, such as by monitoring the carotid artery, is particularly impactful for preventive care. Real-time imaging and monitoring in such devices improve diagnostic outcomes and patient management in both clinical and telemedicine settings~\cite{recker2022ultrasound}.
Ensuring timely and quality healthcare in these regions requires targeted efforts to improve infrastructure and resource availability.


However, transmitting medical images from rural areas to city hospitals or cloud services presents several obstacles. Limited bandwidth and unreliable networks hinder the quick transfer of large files (e.g., medical images and/or short video), which impedes timely and effective medical consultations and interventions. Additionally, rural mobile networks often lack sufficient coverage and bandwidth, complicating the transmission of media-heavy content. Moreover, IoT devices used for these transmissions typically have restricted battery life, with scarce access to charging. The sparse infrastructure, few base stations, and resource-sharing among operators further exacerbate these challenges.
Data compression becomes essential in this context, allowing for more efficient communication over the wireless link. 
Several approaches have been proposed to reduce the size of the transmitted data over the wireless network including semantic communication~\cite{Ismail_2022_FNWF}, source coding~\cite{Golmohammadi_2022_TCOM} and compressive sensing for ultrasound imaging~\cite{liu2016compressed}.
Recent advancements in machine learning, particularly Autoencoders (AEs), provide a promising solution for data compression~\cite{vepakomma2018split}. AEs are capable of encoding high-dimensional data into compact, latent representations, thereby reducing the communication overhead without sacrificing essential information. 
For example, AEs have been used in video streaming applications to reduce bandwidth requirements while maintaining video quality~\cite{habibian_2019_video}.
This is expected to allow more users to obtain channels and users with low latency requirements can deliver their data faster.


Another critical issue for telemedicine in rural area is the efficiency of the deployed spectrum allocation mechanisms where both the availability and demand for spectrum differ significantly from urban environments.
Traditional spectrum allocation methods, such as fixed pricing or administrative assignments, are ill-suited to handle the variable demand in these sparsely populated regions. 
Fixed term contracts and prices are not beneficial for both the spectrum service provider (SSP) and the mobile users as they lack adaptability to fluctuating demand. Specifically, for the SSP, some mobile devices might be ready to pay higher than the average fixed price values and hence can increase his revenue if a dynamic pricing mechanism is used.
For example, users that want to use the spectrum for entertainment might not be willing to pay high as a user which has important data to transmit urgently, e.g., an on-site medical worker.
From the other side, the mobile user might be inactive at certain periods of time and hence paying for resources that is not using.
Advanced spectrum allocation frameworks that can optimize resource use in rural settings while ensuring fairness, accessibility and maximize revenue are therefore required. 
Auction theory has emerged as a powerful tool for spectrum resource allocation, offering a market-driven approach where users bid based on their needs and willingness to pay~\cite{Yang_2013_Auction_survey, Ismail_2022_FNWF}. This ensures that spectrum is allocated to those who value it most, resulting in more efficient resource use compared to fixed allocation methods. 
In a scenario where spectrum demand varies significantly such as between users streaming entertainment content and users transmitting critical data (e.g., a medical worker), the auction mechanism efficiently prioritizes the latter by letting users bid based on their urgency and willingness to pay. This ensures that essential services with higher urgency and utility are prioritized, promoting the effective use of spectrum.






\subsection{Contribution}
While AEs have been successful in compressing data for individual users, their integration with auction mechanisms—where multiple users compete for spectrum—is still not well investigated. Most studies focus on standalone compression tasks without considering the joint optimization of spectrum allocation and communication efficiency, e.g., as in~\cite{habibian_2019_video, recker2022ultrasound}.
This paper proposes a hybrid framework that combines auction theory for spectrum resource trading with AEs for telemedicine data compression over wireless links. 
This joint optimization of resource allocation and data compression enhances the overall efficiency of the rural wireless network, improving connectivity for users and maximizing the social welfare of the system.
To the best of our knowledge, this is the first paper that study the spectrum auction market for users with AE-based data compression.
The main contributions of this work are as follows:
\begin{itemize}
    \item First, we introduce an auction-based spectrum allocation mechanism tailored for rural areas, focusing on incentivizing participation and maximizing social welfare. The system ensures that these high-urgency users not only win the necessary spectrum but also transmit their data more quickly, optimizing the use of the spectrum resource.

    \item Second, we integrate AEs into the framework for efficient data compression, significantly reducing communication overhead in bandwidth-constrained environments. While auction theory ensures that spectrum goes to the users who need it most, AEs enhance this process by allowing more users to share the available spectrum through data compression. Furthermore, the use AE helps transmitting the data securely over the wireless link as only the receive can decode the transmitted latent data.

    \item Finally, we validate the proposed framework through extensive simulations, demonstrating its superior performance in rural scenarios compared to traditional spectrum allocation methods. The results show that our approach not only improves spectrum utilization and communication efficiency but also enhances connectivity and fairness in rural wireless networks.
\end{itemize}

In Section~\ref{sec:system_model}, we describe our proposed AE-based marketing model. In Section~\ref{section:SW_max} we formulate the social welfare maximization problem. 
Simulation results are presented in Section~\ref{section:simulation} and Section~\ref{section:conclusion} concludes the paper.

\section{System Model}\label{sec:system_model}
Figure~\ref{fig:system_model_conf} describes our proposed system design tailored for  real-time medical assistance scenarios. 
Note here that other variant use cases can be obtained with slight modifications to Figure~\ref{fig:system_model_conf}. For instance, in the case of wireless sensors transmitting ultrasound images for remote health monitoring, immediate feedback from hospitals may not be necessary for every transmitted data. However, the core problem under study is the uplink allocation of spectrum resources and how the data size can be minimized.
In the following discussion, we refer to the real-time medical assistance use case to illustrate the practical application of our proposed solution.

\subsection{System Overview}
\begin{figure}[ht!]
    \centering    \includegraphics[width=.4\textwidth,height=4.cm]{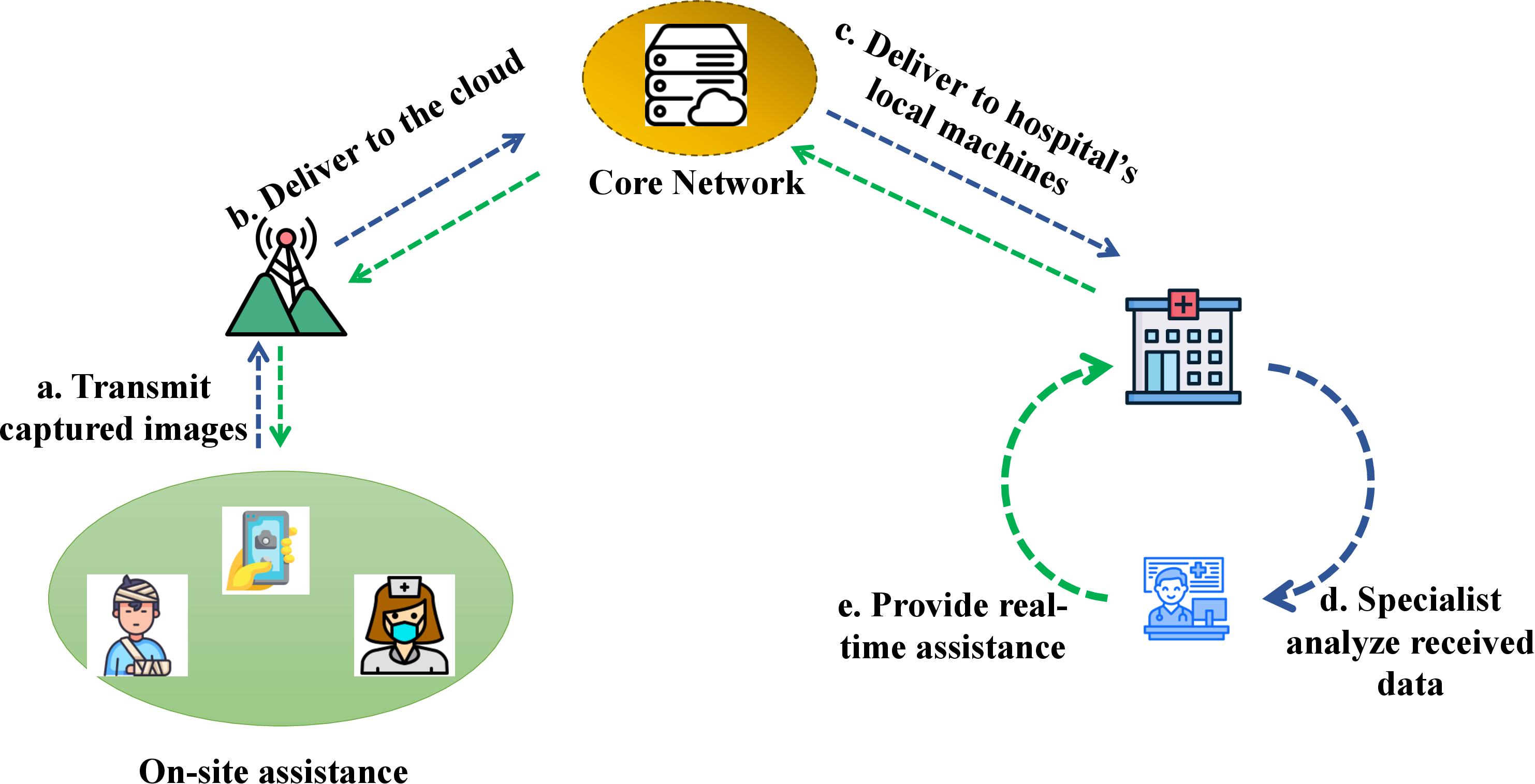}
    \caption{Proposed system model for real-time telemedicine assistance in rural areas.}
    \label{fig:system_model_conf}
\end{figure}

As illustrated in Figure~\ref{fig:system_model_conf} , the on-site medical health worker first captures a set of images of the patient using available medical screening tools (e.g., x-ray images) or phone-based images and videos (e.g., for a patient with bleeding or suspected broken bones). Next, the collected images are delivered to the base station (BS) with highest signal-to-noise ration (SNR) (step a in Figure~\ref{fig:system_model_conf}). The BS then delivers the content to the cloud servers (step b in Figure~\ref{fig:system_model_conf}), which forward the content to the hospital's local machines (step c in Figure~\ref{fig:system_model_conf}). Once the medical images and their metadata are received by the hospital, doctors start analyzing the patient's situation (step d in Figure~\ref{fig:system_model_conf}) and then send back their advice and assistance (step e in Figure~\ref{fig:system_model_conf}) to the on-site medical healthcare workers to take necessary actions before the patient is transported to the hospital.

However, as the patient and the on-site health workers are located in a rural area, the quality of their connected wireless devices (e.g., mobile phones) to the BS is likely to be low, which poses delays to the remote medical assistance. Specifically, limited network infrastructure such as fewer cell towers and greater distance from cell towers lead to weaker SNR between the mobile device and the BS. Additionally, numerous users (with different preferences) competing for access to the same spectrum further congest the network. 
Furthermore, as media content is large in volume size, longer delays are expected to have the full set of content uploaded to the cloud servers and received by the hospital's doctors.

To address the aforementioned challenges, we propose utilizing \acp{AE} to compress the size of transmitted medical data, significantly reducing the communication overhead. Additionally, an auction theory framework is introduced to optimize spectrum allocation, ensuring that users who value the bandwidth the most receive priority. This dual approach aims to enhance both the efficiency of data transmission and the fair distribution of limited wireless resources in rural settings, improving connectivity and response times for real-time medical assistance.

\subsection{Autoencoder-Based Data Compression}








To reduce data size, we consider the combination of two techniques: \acp{AE} and \ac{IB}. \acp{AE} are generative models that learn latent representations of input data, which can be used for efficient encoding and decoding. The key components of a AE include an encoder, a decoder, and a latent space representation. AEs have been successfully applied to various image processing tasks, including denoising, super-resolution, and compression~\cite{Geiger_2021_TNNLS}.
To decide about how much information is still prserved withing the latent space, \ac{IB} is used. \ac{IB} theory focuses on finding a trade-off between the amount of information retained about the input data and the complexity of the representation. In the context of image transmission, IB theory helps in identifying and preserving the most relevant features for diagnosis while compressing the data.

\begin{figure}[ht!]
    \centering    \includegraphics[width=.4\textwidth,height=4.cm]{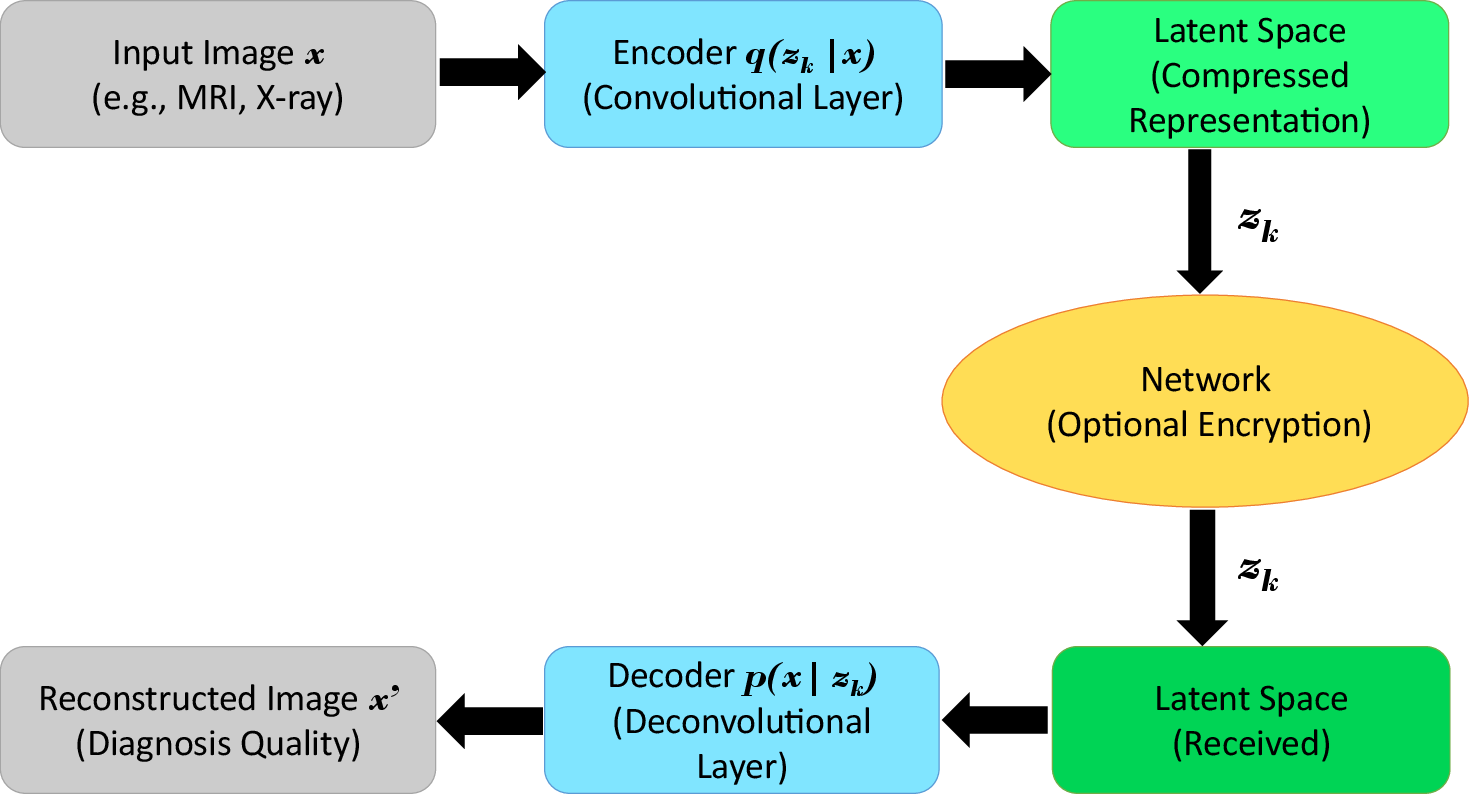}
    \caption{Overview of a general AE-based medical image compression inference pipeline.}
    \label{fig:system_autoencoder}
\end{figure}

Figure~\ref{fig:system_autoencoder} describes the overflow of an image transmitted over the wireless link with AE compression. 
The process starts with a medical image, such as an MRI or X-ray. The input image is passed through the encoder, which consists of several convolutional layers. These layers reduce the dimensionality of the image while preserving essential features.
The encoder then outputs a compressed representation of the image, known as the latent space. This representation is significantly smaller in size compared to the original image.
The latent space representation is transmitted over a wireless network. For enhanced security, this data can be encrypted to ensure privacy and protect sensitive patient information\footnote{Note here that there is no need for encrypting the data as the latent representation can be recovered only by the receiver (unless brute-force techniques are used). Nevertheless, AE can be further trained on encrypted data.}.
The received latent space is passed through the decoder, which consists of deconvolutional layers. These layers reconstruct the original image from its compressed representation.
Finally, the reconstructed image is expected to retain the quality necessary for accurate medical diagnosis.
This architecture helps address several challenges by providing efficient compression, maintaining image quality, and ensuring secure transmission of medical images over wireless networks.

Note here that several works have addressed the objective of optimizing the trade-off between data compression and the informativeness of the recovered data, e.g.,~\cite{Omar_Globecom_2023, Lungisani_2023_Access}. However, that is not the objective of our work. Instead, our system capture the use of the \ac{IB} theory and encoder-decoder architecture as a general function that takes as input the data to compress and output the informativeness score that can be achieved in addition to the new data size of the compressed data (i.e., latent space). These information are then used in the utility function of the \acp{UE} to decide about the number of channels to request and the bid value to submit, which we detail in what follows.

\subsection{Valuation Metrics}
Each IoT device needs to calculate their valuation of the spectrum and the number of channels to request from the SSP. 
The IoT devices first set their latency requirements $T=\{T_1, T_2, \dots, T_i, \dots, T_N\}$ where $T_i$ is the latency of IoT device $i$ and $N$ is the total number of IoT devices participating the auction market. 
The latency is mainly captured by the uplink transmission time from the IoT device to the nearest base station and is defined as
\begin{equation}
    T_i = \frac{\hat{d_i}}{r C_i},
\end{equation}

\noindent where $d_i$, $\hat{d_i}$, $C_i$ and $r$  are data size before applying the autoencoder, data size after applying the autoencoder, the number of channel allocated to IoT device $i$ and the channel capacity, respectively.
The channel capacity at the receiving base station is defined as
\begin{equation}
    r = M \log(1+\frac{g P_{Trans}}{N_0^2}),
\end{equation}

\noindent where $g$ is the channel gain, $P_{Trans}$ is the IoT device transmit power and $N_0^2$ is the noise power spectral density. $M$ is the channel bandwidth in $Hz$.


We consider that each IoT device has a data queue for the generated media content (e.g., medical images) to be sent to the base station. This queue temporarily holds incoming data packets before transmission. The queue size is represented by the vector $\mathbf{q}=( q_{1}, q_{2}, \dots, q_{N})$. For IoT device $i$, the queue is filled at a constant rate  $\frac{1}{\lambda}$, modeled by a Poisson distribution, as in~\cite{Ismail_2021_Globecomm}. Let $P_i(q_i)$ be the probability that IoT device $i$ performs a successful data transmission.
Since spectrum allocation is fixed over time, the probability of transmission failure increases with queue size. The probability of failure is approximated using a Poisson distribution, expressed as:
\begin{equation}\label{eq:someequation}
    \mathcal{P}_i^f =  1 - \exp(-\frac{1}{\lambda}T_i), 
\end{equation}

Hence, the probability of successful transmission for IoT device $i$ is given by:

\begin{equation}\label{eq:someequationwwwwwww}
\begin{multlined}
    \mathcal{P}_i(q_i) =  1 - \mathcal{P}_i^f 
    = \exp(-\frac{1}{\lambda} T_i).
\end{multlined}
\end{equation}

Therefore, the valuation of IoT device $i$ towards $C_i$ channels is formulated as a discounted value based on the probability of successful transmission, which defined as
\begin{equation}\label{eq:vi_C_i}
    v_i^{C_i} = C_i \exp(-\frac{1}{\lambda} T_i)
\end{equation}

Additionally, the benefits from using the autoencoder compressor is formulated as
\begin{equation}\label{eq:vi_AE}
   v_i^{AE} = w_i^1 (d_i - \hat{d_i}) + w_i^2 (T'_i - T_i) + w_i^3 (Acc_i),
\end{equation}

\noindent where $T'_i$ is the latency for the case where no compression is used, and $w_i^1$, $w_i^2$ and $w_i^3$ are weighting factors with $w_i^1+w_i^2+w_i^3=1$. $Acc_i$ is the accuracy of the data retrieved by the decoder. The weighting factors reflects the subjective importance of compressed data size, latency and reduced accuracy due to the use of \acp{AE}.

By summing-up \eqref{eq:vi_C_i} and \eqref{eq:vi_AE}, the total valuation of IoT device $i$ towards using the spectrum channels is defined as
\begin{equation}\label{eq:v_i}
\begin{multlined}
    v_i = v_i^{C_i} + v_i^{AE} \\
    = C_i \exp(-\frac{1}{\lambda} T_i) + w_i^1 (d_i - \hat{d_i})\\ + w_i^2 (T'_i - T_i) + w_i^3 (Acc_i).
\end{multlined}
\end{equation}


Finally, each IoT device is also required to compute the cost required to perform the \ac{AE}-based compression and the transmission cost.
The computation cost for using the autoencoder compression technique by IoT device $i$ is denoted as $c_i^{AE}$ while the communication cost is defined as~\cite{Yutao_2020_TMC}
\begin{equation}\label{eq:c_i_m}
    c_i^{Trans} = r C_i \beta_i ,
\end{equation}

\noindent where $\beta_i$ is the per unit energy cost for transmission. The total service cost is then defined as
\begin{equation}
    c_i = c_i^{Trans} + c_i^{AE}
\end{equation}


\subsection{Auction-Based Spectrum Allocation Mechanism}
In our considered system model, a multi-buyer single single item auction mechanism to formulate the problem. Although the IoT devices (the buyers) can request for multiple channels, they are also single minded, i.e., each device either wins all the channels it requests or none at all. This property simplifies the bidding strategy for participants, as they do not receive partial allocations~\cite{Nisan_2007}. 
Each IoT device bids for the price it is willing to buy the spectrum resources from the SSP. This model incentivize the IoT devices to strategically increase their bids $\mathbf{b} =( b_{1}, b_{2}, \dots, b_{N})$ to secure access to the minimum number of channels necessary for transmitting their data efficiently. The number of channels requested by each IoT device is formulated as profile vector $\mathbf{C}=( C_1, C_2, \dots, C_i, \dots, C_N)$.
After receiving the bids and demands from IoT devices, the SSP determines the winners of the auction and informs all participating IoT devices about their allocation \(\mathbf{\psi} = (\psi_1, \psi_2, \dots, \psi_N)\) and the corresponding spectrum prices \(\mathbf{p} = (p_1, p_2, \dots, p_N)\). Here, \(\psi_i = 1\) indicates that user \(i\) is a successful bidder and has been allocated channels, while \(\psi_i = 0\) means that user \(i\) lost the auction. \(p_i\) represents the price user \(i\) must pay for their allocated spectrum.

Our auction design aims to achieve three key properties: (i)~\textbf{Truthfulness}, where bidders maximize their utility by honestly declaring their true valuation and information, regardless of others' bids;  (ii)~\textbf{Individual rationality}, guaranteeing non-negative utility for each bidder and a positive net profit for the SSP; and (iii)~\textbf{Social welfare maximization}, optimizing the overall efficiency of the wireless network system through truthful bidding.











\section{Social Welfare Maximization Auction}\label{section:SW_max}

The utility of an IoT device $i$ is the difference between its valuation of the spectrum and its payment $p_i$ and service cost $c_i$, and is expressed as
\begin{equation}\label{eq:u_i}
    u_i = v_i - p_i - c_i.
\end{equation}

The utility of the SSP is defined as the total payment received from all the winning IoT devices and the service cost $\hat{c}$, e.g., in terms of energy consumption while processing the data, and is expressed as
\begin{equation}\label{eq:u_SSP}
    \hat{u} = \sum_{i\in N} \psi_i p_i   - \hat{c},
\end{equation}

\noindent where $\psi_i$ is a binary variable that indicates if an IoT device $i$ is amongst the winners, i.e., $\psi_i=1$, or not amongst the winners, i.e., $\psi_i=0$.
The social welfare of the system is defined as the sum of the utilities of all participating entities, including both the SSP and the IoT devices. Mathematically, the total social welfare is expressed as:
\begin{equation}
\begin{multlined}
    SW(\psi) = \sum_{i\in N} \psi_i u_i  + \hat{u}\\
    = \sum_{i\in N} \psi_i (v_i-p_i-c_i) + \sum_{i\in N} \psi_i p_i + \hat{c}\\
    = \sum_{i\in N} \psi_i (v_i-c_i)  + \hat{c}\\
\end{multlined}
\end{equation}

The social welfare maximization problem can be formally written as an integer linear programming (ILP) problem, aiming to optimize the overall system efficiency~\cite{Zhang_2017_welfare}. The objective is to maximize the total social welfare, which includes the utilities of both IoT devices and the SSP, subject to spectrum availability, which we formulate as

\begin{subequations}
\label{eq:optz_1}
\begin{align}
\begin{split}
\max_{\psi} SW(\psi) = \sum_{i\in N} \psi_i (v_i-c_i)  + \hat{c}, \label{eq:MaxA} 
\end{split}\\
\begin{split}
\hspace{1cm} s.t. \sum_{i\in \mathcal{N}} \psi_i C_i \leq B\label{eq:MaxB}
\end{split}\\
\begin{split}
\hspace{1cm} \psi_i \in \{0,1\}, \forall i\in \mathcal{N} \label{eq:MaxC}
\end{split}
\end{align}
\end{subequations}

\noindent where $B$ is the total number of channels provided by the SSP.
To ensure a fair and robust market, the payment mechanism for winning IoT devices must satisfy the properties of \ac{IC} and \ac{IR}. \ac{IC} ensures that each IoT device gains the highest utility by truthfully bidding based on its true valuation. This discourages manipulation of bids to achieve undeserved higher utility. \ac{IR} ensures that no IoT device incurs a negative utility, meaning that the device will always prefer participating in the auction to not participating.
\textcolor{black}{In what follows, we present the payment rule for winning IoT devices and prove the properties of IC and IR.}

\subsubsection{Payment Rule}
The payment rule for the winning IoT devices is based on the Vickrey-Clarke-Groves (VCG) auction mechanism. In this mechanism, each winning device pays an amount equal to the harm it causes to other users by taking the spectrum resource, rather than paying its bid~\cite{Zhang_2017_welfare} . This ensures that devices bid truthfully (IC). Formally, the payment rule is represented as
\begin{equation}\label{eq:payment}
    p_k = SW(\psi^*) - SW_{\mathcal{N}\backslash\{k\}}(\varphi^*),
\end{equation}

\noindent where $\psi^*$ represents the optimal spectrum allocation strategy for all participating IoT devices, based on their bids and demand vectors, and $S(\psi^*)$ denotes the maximum social welfare attained under this allocation. On the other hand, $S_{\mathcal{N}\backslash\{k\}}(\varphi^*)$ is the maximum social welfare when IoT device $k$ is excluded from the auction, where $\varphi^*$ is the optimal allocation for the remaining devices.

\subsubsection{Incentive Compatibility and Individual Rationality}

\begin{definition}[Incentive Compatibility]
    An IoT device $i$ has no incentive to misreport its true valuation $v$.  This is because for any other false bid $v'$, the utility it would receive is lower than the utility it gains by submitting $v$.
    Formally,
    {\small \begin{equation}\label{eq:IC_property}
        v'_{i}-p^{(v_i')}_i \leq v_{i} -p^{(v_i)}_i, \quad \forall i\in \mathcal{N},
    \end{equation}}\noindent
    where $p^{(v_i)}_i$ and $p^{(v_i')}_i$ are the obtained payments for the true valuation $v$ and any other valuation $v'$, respectively.
    
\end{definition}

\begin{definition}[Individual Rationality]
    The utilities for all the IoT devices must be non-negative, i.e., $u_i \geq 0, \forall i \in \mathcal{N}$.
\end{definition}

\begin{theorem}\label{them_1}
	The proposed VCG-based auction mechanism is both incentive compatible and individually rational.
\end{theorem}
\begin{IEEEproof}
    Since the payment rule in \eqref{eq:payment} follows the VCG condition, the proof follows from the one derived in~\cite{Ismail_2022_FNWF}.

\end{IEEEproof}

\section{Performance Evaluation}\label{section:experiments}\label{section:simulation}

In this section, we present a numerical evaluation of our proposed hybrid framework for medical data compression and spectrum allocation in rural areas. Our analysis focuses on key performance metrics such as spectrum efficiency, latency reduction, and system scalability. 
We compare our proposed auction algorithm with the clearing auction scheme, where the price is determined by the lowest bid among the winners and all winning bidders are required to pay at least the minimum winning bid. This mechanism incentivizes bidders to bid just high enough to win, as bidding lower risks losing the auction, but overbidding leads to unnecessary costs. 
We also compare two scenarios: one where AE-based data compression is employed and another where it is not. We analyse how compression affects spectrum allocation and the utility of IoT devices. 
Note that we use \emph{Gurobi optimizer} to solve the ILP in \eqref{eq:optz_1}.

\subsection{Impact of the AE-based compression on the winner list and system efficiency}
\begin{figure}[ht!]
    \centering
    \includegraphics[width=.4\textwidth,height=4.cm]{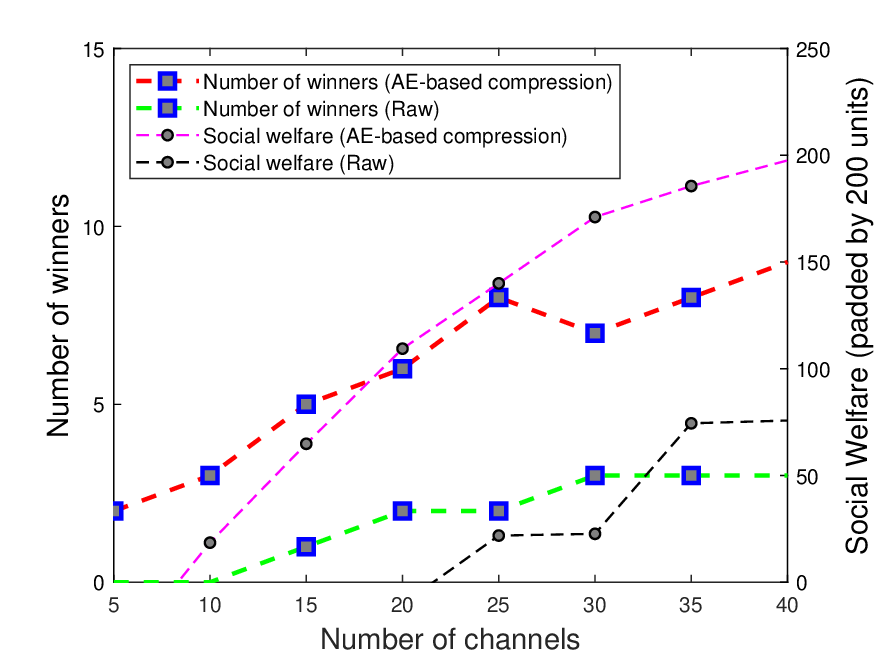}
    \caption{Impact of the number of channels on the number of winners and social welfare.}
    \label{fig:conf_result_1}
\end{figure}

Here, we evaluate two transmission types: the first one consists of transmitting raw data, i.e., no AE-based compression, and the second one consists of transmitting the latent space provided by the AE, i.e., AE-based compressed data.
From Figure~\ref{fig:conf_result_1}, we see that as the number of channels \( B \) increases, the number of winners in the AE-compression case grows significantly, reaching up to 9 winners, while for raw data transmission (i.e., no AE-based compression), the winners remain low (1-2). The larger size of raw data restricts spectrum usage. Interestingly, when \( B \) increases from 25 to 30, fewer devices win due to the higher value of the data from selected devices, leading to greater social welfare despite fewer winners.
We observe that with compression, the reduced data size allows for more efficient spectrum usage, enabling more devices to transmit within the same bandwidth. This leads to increased utility for IoT devices due to faster transmission times and lower communication costs. In contrast, without compression, devices experience higher spectrum contention and reduced overall utility.

\subsection{Impact of the truthfulness of the auction mechanism}
\begin{figure}[ht!]
    \centering
    \includegraphics[width=.4\textwidth,height=4.cm]{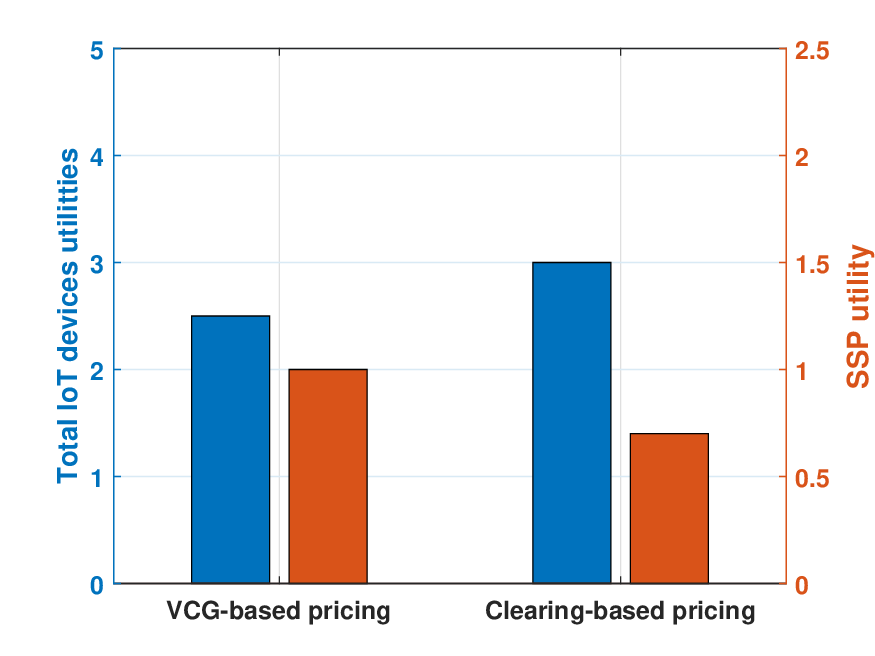}
    \caption{Impact of the payment rule on the utilities of the SSP and the IoT devices.}
    \label{fig:conf_result_2}
\end{figure}

To capture the importance of the truthfulness (i.e., IC) property of our VCG-based auction mechanism, we consider the use of the clearing auction scheme and plot the results in Figure\ref{fig:conf_result_2}.
In clearing auctions~\cite{derksen2020clearing}, all winners are considered to pay the same price and is equal to the highest losing bid.
Although the payment rule for clearing auction ensures fairness by making all winners pay the same amount, it is not incentive compatible.
In such auctions, bidders may have an incentive to underbid in order to pay a lower price.
We observe from Figure\ref{fig:conf_result_2}, that the total utility for winning IoT devices in the clearing auction is higher than that in the VCG-based auctions. Conversely, the utility of the SSP in the clearing auction is lower than that in the VCG-based auctions. 
This is justified by the fact that as the optimization problem in\eqref{eq:optz_1} is independent of the payment rule, we will have the same set of winners, however, as the payment rule changes, the total utility of the IoT devices and the SSP will change
Specifically, as the prices in the clearing auction are lower than those in the VCG-based auction, the utility of the IoT devices increases as defined in\eqref{eq:u_i}.
Similarly, the utility of the SSP in the clearing auction is much lower due to the received low payments, as defined in~\eqref{eq:u_SSP}.

\subsection{Discussion}
Our proposed framework serves as a theoretical model for investigating how market-based mechanisms could enhance resource utilization. Rather than replacing current allocation systems like OFDMA, we propose integrating auction-based prioritization to complement them, particularly in high-demand situations such as emergencies. The auction mechanism could be implemented within the operator's scheduling layer, working alongside existing allocation methods to enable prioritization while maintaining user contracts. However, we acknowledge that this integration introduces computational overhead that requires careful evaluation in terms of system scalability.

\section{Conclusion}\label{section:conclusion}


This paper introduced a hybrid framework that combines AEs for data compression with auction-based spectrum allocation to optimize both communication efficiency and spectrum utilization for remote healthcare systems. Auction theory allocates spectrum based on need and willingness to pay, while AEs improve the overall efficiency of the system by compressing data, to allow more users to participate. Together, they create a more dynamic, efficient, and equitable spectrum allocation system, especially in scenarios with varied user requirements and limited bandwidth.
By reducing the data overhead and dynamically allocating resources based on demand, this framework enhances connectivity, improves healthcare delivery in remote areas, and facilitates timely transmission of medical data. 
Implementing the framework on real datasets is a key next step in our future work.








\bibliographystyle{IEEEtran}
\bibliography{reference}

\end{document}